# High frequency magnetic permeability of nanocomposite film

N.A. Buznikov, I.T. Iakubov, A.L. Rakhmanov, and A.O. Sboychakov[*]

*Institute for Theoretical and Applied Electromagnetics, Russian Academy of Sciences, Moscow 125412, Russia*

**Abstract**

The high frequency magnetic permeability of nanocomposite film consisting of the single-domain spherical ferromagnetic particles in the dielectric matrix is studied. The permeability is assumed to be determined by rotation of the ferromagnetic inclusion magnetic moments around equilibrium direction in AC magnetic field. The composite is modeled by a cubic array of ferromagnetic particles. The magnetic permeability tensor is calculated by solving the Landau-Lifshits-Gilbert equation accounting for the dipole interaction of magnetic particles. The permeability tensor components are found as functions of the frequency, temperature, ferromagnetic inclusions density and magnetic anisotropy. The obtained results show that nanocomposite films could have rather high value of magnetic permeability in the microwave range.

## 1. Introduction

The magnetic nanocomposites are intensively studied. The interest to the physical properties of these materials arose more than fifty years ago and now is renewed due to possibility of their application in magnetic field sensors and magnetic recording systems. The composites consisting of iron, cobalt or nickel ferromagnetic single-domain nanoparticles in metal, semi-conducting or insulating matrix demonstrate extraordinary magnetic and transport properties (superparamagnetism, giant magnetoresistance, tunneling magnetoresistance, etc). The nanocomposites with insulating matrix are quite promising for application at high frequencies. The large value of permeability has been demonstrated for composite thin films with cobalt nanoparticles within the frequency range up to 1 GHz [1-3]. The films structure constitutes an array of non-oxidized cobalt particles in insulating (e.g., zirconium oxide) matrix. The cobalt particles with the characteristic size of several nanometers had nearly spherical shape and rather narrow distribution in diameters. Due to relatively regular structure, such composites are insulators at ferromagnetic metal content significantly higher than that for the random systems. Note that the intriguing high-frequency properties of studied magnetic nanocomposites are determined by the large magnetization and magnetic anisotropy of Co nanoparticles [4-6] as well as a relatively high (of thousands $\mu\Omega\cdot$cm) electric resistance of the material before the insulator – metal percolation transition.

The magnetic structure of nanocomposites is a separate and rather exciting problem. It is a subject of a set of theoretical and experimental investigations [1,6-9]. However, the impact of the composite magnetic structure on their high frequency permeability has not been studied in detail till now. It should be also noted that the films studied in the papers [1-3] have ferromagnetic phase content higher than the percolation threshold and, consequently, the magnetic ordering in them arises due to exchange interaction.

---

[*] Corresponding author. Institute for Theoretical and Applied Electromagnetics RAS, Izhorskaya str., 13/19, Moscow 125412, Russia; Tel: +7-095-3625147; e-mail: sboycha@mail.ru

The present theoretical study is devoted to the high frequency permeability of magnetic nanocomposite films. It is assumed that the material consists of the single-domain ferromagnetic metal nanoparticles in nonmagnetic insulating matrix. The ferromagnetic particle volume content, $p$, is supposed below the percolation threshold $p_c$ and the film is in insulating state. We assume also that the magnetic granules are spherical and have uniaxial magnetic anisotropy. It was mentioned before that the percolation threshold in nanocomposite reaches rather high values ($p_c \sim 0.5$). It is due to the existing of amorphous dielectric shell around magnetic granules [1,4,6,10].

There exist two deferent mechanisms governed the nanocomposite permeability. The first mechanism, so-called relaxation mechanism, is related to the process of spontaneous (thermal) reorientation of nanoparticles magnetic moments from one equilibrium position to another [11,12]. Such a process dominates at relatively high temperatures and low frequencies for composites consisting of small ferromagnetic particles with low value of magnetic anisotropy constant. The second mechanism, which is considered herein, is related to the rotation of the ferromagnetic particles magnetic moments near the equilibrium position in the AC magnetic field. The latter process dominates in the case of nanocomposites with magnetic grains of not too small sizes. To find the criterion of applicability of the second mechanism, one should to compare the characteristic magnetic anisotropy energy of the ferromagnetic nanoparticle $KV_f$ with the thermal energy $k_B T$, where here $K$ is the magnetic anisotropy constant, $V_f$ is the particle volume, $k_B$ is the Boltsman's constant, and $T$ is the temperature. Below it is assumed that $KV_f \geq k_B T$. The last inequality determines the minimal particle diameter

$$D \geq D_{\min} = (12 k_B T / \pi H_a I_s)^{1/3} , \qquad (1)$$

where $I_s$ is the saturation magnetization of the ferromagnetic material and $H_a = 2K/I_s$ is the anisotropy field. Taking for estimates the parameter values characteristic for Co, $I_s$ = 1500 Gs and $H_a$ = 100 Oe, one finds $D_{min}$ = 10 nm at $T$ = 300 K. Let us introduce dimensionless parameter $\lambda = KV_f / k_B T$. If $\lambda \geq 1$, that is $D \geq D_{min}$, then the reorientation time of magnetic moment due to thermal fluctuations can be estimated as [11,12]

$$\tau_{rel}^{-1} = \gamma H_a \sqrt{\lambda / \pi} \exp(-\lambda) , \qquad (2)$$

where $\gamma$ is the gyromagnetic ratio. The contribution of the magnetic moment thermal relaxation to magnetic permeability is evidently small if the AC magnetic field frequency $f$ is large, $f > f_{min} = 1/2\pi\tau_{rel}$. For example, at $\lambda = 1$ and $H_a$ = 100 Oe we get $f_{min} \approx$ 1 GHz. We assume below that $D > D_{min}$ and $f > f_{min}$.

The properties of the composite depend strongly on the value of the magnetic interaction between ferromagnetic granules. This interaction can arise due to both exchange and dipole forces. The former is of importance if the characteristic space between the granules, $\Delta r$, does not exceed significantly the exchange depth, $l_{ex}$, which is of the order of 1 nm. The intergranular distance can be estimated as $\Delta r = D(p^{-1/3} - 1)$. In the present paper it is assumed that $\Delta r \gg l_{ex}$. It means, in particular, that the composite is in the insulating state and not close to the percolation threshold over ferromagnetic inclusions and the granule sizes is not too small, e.g., D = 10 nm. In such a case the interaction between granules is mainly due to the dipole contribution. The dipole interaction could be characterized by the ratio of characteristic dipole and thermal energies, $\eta = 8 p V_f^2 I_s^2 (D^3 k_B T)^{-1}$. The present study is valid under condition of a strong dipole interaction, that is, at $\eta \gg 1$.

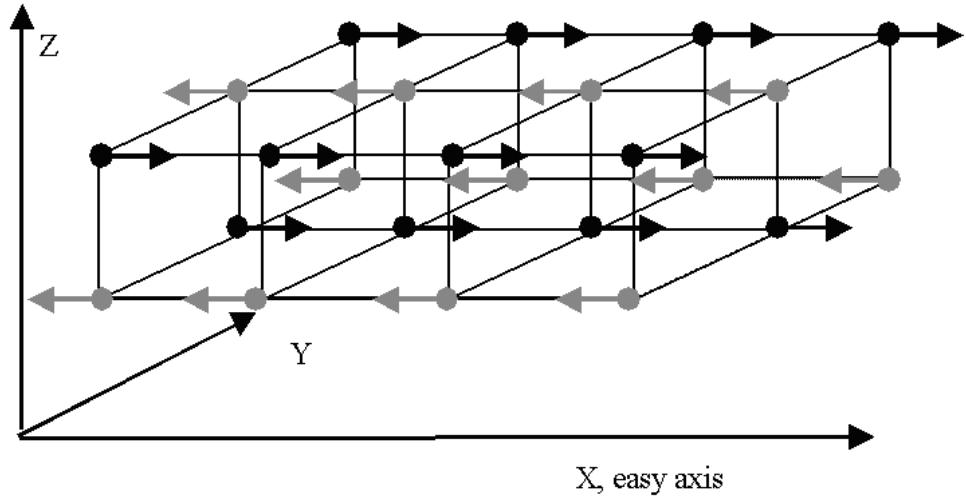

**Fig. 1.** Schematic of the antiferromagnetic ordering of the nanoparticles magnetic moments.

To calculate magnetic permeability of a composite we should find equilibrium orientations of the particles magnetic moments accounting for the magnetic interaction between them and then determine a variation of this energy at small oscillations of the magnetic moments near their equilibrium positions. In general, it is a difficult task to find a solution to this problem even numerically in the case of disordered composite with not too small magnetic particles content when the interaction between magnetic inclusions is strong. Following the approach developed in [13, 14], we shall model the composite by ordered array of magnetic particles in the insulating matrix. To simplify the problem, we assume that the magnetic inclusions constitutes a cubic lattice with the axes $x$, $y$, and $z$ along the cube edges, $z$-axis is directed across film surface and $x$, $y$ directions lie in the film plane. We also assume that all the particles have the same diameter $D > D_{min}$, their magnetic easy axes are parallel and directed in the film plane (along $x$-axis). Note that, in principal, such a distribution of anisotropy axes could be attained using special film deposition regimes, e.g., under deposition in the applied DC magnetic field.

## 2. Energy of dipole interaction

The energy of the magnetic dipole interaction between spherical particles is equal to

$$U_{int} = \frac{1}{2}\sum_{i,j}{}' \left( \frac{\mathbf{m}_i \mathbf{m}_j}{|\mathbf{r}_i - \mathbf{r}_j|^3} - \frac{3(\mathbf{m}_i(\mathbf{r}_i - \mathbf{r}_j))(\mathbf{m}_j(\mathbf{r}_i - \mathbf{r}_j))}{|\mathbf{r}_i - \mathbf{r}_j|^5} \right), \qquad (3)$$

where $\mathbf{r}_i$ is the vector connecting the coordinate origin with the center of $i$-th particle. It is known that the antiferromagnetic configuration of C-type corresponds to the energy minimum of such a system at zero temperature [15,16]. That is, in the equilibrium state the magnetic moments of particles directed along the easy axis making up two sub-lattices with antiparallel moments (see Fig. 1).

We can write the magnetizations of two sub-lattices as

$$\mathbf{M}_k = (pI_s/2)\mathbf{e}_k, \qquad k = 1,2, \qquad (4)$$

where $\mathbf{e}_k$ are the unit vectors determining the magnetization direction for each sub-lattice. Then, the magnetic energy of the composite system per unit volume can be presented in the form

$$u = \mathbf{M}_1 \Lambda_{AF} \mathbf{M}_2 + \frac{1}{2} \sum_{k=1,2} \mathbf{M}_k \Lambda_F \mathbf{M}_k - \frac{H_a}{pI_s} \sum_{k=1,2} (M_{k,x})^2 , \qquad (5)$$

where $\Lambda_F$ and $\Lambda_{AF}$ are the tensors determining the values of particles interaction in one sub-lattice and in different sub-lattices correspondingly. The components of the tensors are defined by expression (3), and in the coordinate system connected with the cubic lattice axes they can be presented as following sums,

$$(\Lambda_{F,AF})_{\alpha\beta} = \sum_{n_x=-\infty}^{\infty} \sum_{n_y=-\infty}^{\infty} \sum_{n_z=1-k}^{Z-k} (1 \pm (-1)^{n_y + n_z}) \times$$
$$\times \left( \frac{\delta_{\alpha\beta}}{[(n_x)^2 + (n_y)^2 + (n_z)^2]^{3/2}} - \frac{3 n_\alpha n_\beta}{[(n_x)^2 + (n_y)^2 + (n_z)^2]^{5/2}} \right), \qquad (6)$$

where signs of plus and minus refer to the components of tensors $\Lambda_F$ and $\Lambda_{AF}$ respectively; $Z$ is the number of layers in the composite film, and $k = 1, 2, \ldots, Z$ is the layer number for which interaction energy is calculated. Numerical calculations of tensor components $\Lambda_F$ and $\Lambda_{AF}$ were performed for films containing $Z = 7$ and $Z = 101$ layers. The film thickness is correspondingly 0.1 μm and 1 μm if the granule diameter is 10 nm and the granule content is $p = 0.2$. The computations reveal that the values of tensor components vary significantly from layer to layer only near the surface. For the rest layers, the difference is less than one percent. Besides, the values of corresponding tensor components in the central layer for $Z = 7$ and $Z = 101$ are equal within the calculation accuracy. This property is due to both the slow convergence of series (6) and the algebraic structure of terms in them. Having all these in mind, we can assume further that the tensors $\Lambda_F$ and $\Lambda_{AF}$ are the same for all layers and are equal to their values in the central layer. In the chosen above coordinate system ($x,y,z$), only diagonal components of $\Lambda_F$ and $\Lambda_{AF}$ are non-zero and for the central layer we find

$$\begin{cases} \Lambda_{F,xx} = -9.54, \Lambda_{F,yy} = -1.51, \Lambda_{F,zz} = 11.05, \\ \Lambda_{AF,xx} = 1.17, \Lambda_{AF,yy} = -6.87, \Lambda_{AF,zz} = 5.7. \end{cases} \qquad (7)$$

## 3. Magnetic permeability

Now we calculate nanocomposite film permeability in the microwave range. In the considered frequency range $f$, the wavelength of the incident radiation is much greater than the distance between the magnetic particles. Then, one can assume that the magnetic moments of ferromagnetic grains in high frequency magnetic field $\mathbf{H}_0 \exp(-i\omega t)$ oscillates in phase in each sub-lattice. In this case, we can write the Landau-Lifshits-Gilbert equation for magnetization of two sub-lattices in the form,

$$\frac{\partial \mathbf{M}_k}{\partial t} = \gamma \mathbf{M}_k \times \left( \frac{\partial u}{\partial \mathbf{M}_k} - \mathbf{H}_0 \exp(-i\omega t) \right) + \frac{\alpha}{(pI_s/2)} \mathbf{M}_k \times \frac{\partial \mathbf{M}_k}{\partial t} , \qquad (8)$$

where $\alpha$ is the Gilbert damping parameter. The composite permeability tensor $\mu_{\alpha\beta}(\omega)$ can be found by means of a standard procedure of solution of the linearized equations system (8) [17].

It should be noted, however, that expression (5) for the magnetic energy as well as Landau-Lifshits-Gilbert equation in form (8) are valid only at zero temperature. At $T > 0$, one should take into account possible turns over of some particles magnetization vectors due to thermal fluctuations. As a result, some magnetization moments in each sub-lattice will be oriented antiparallel to their equilibrium directions $\mathbf{e}_k$ and the effective magnetic field acting on the particle in $k$-th sub-lattice will differ from that at zero temperature. Those irregular pattern of magnetization we can treat as a quasi-stationary one within the considered temperature $k_B T \ll H_a I_s V_f/2$ and frequency $2\pi f \tau_{rel} \gg 1$ ranges. Following mean-field approach [18], the magnetic energy of the system at $T > 0$ can be found by multiplying of the tensor components $\Lambda_F$ and $\Lambda_{AF}$ in (5) by the factor $\sigma(T)$, which is the ratio of equilibrium magnetization of the magnetic lattice at temperature $T$ to its value at $T = 0$. So, following the standard method [17], one gets for the non-zero components of the magnetic permeability tensor

$$\mu_{xx} = 1,$$
$$\mu_{yy} = 1 + 4\pi\omega_m \frac{\omega_m \Lambda_1 \sigma(T) + \omega_a - i\alpha\omega}{\omega_\parallel^2 - (1+\alpha^2)\omega^2 - 2i\alpha\omega_1\omega}, \qquad (9)$$
$$\mu_{zz} = 1 + 4\pi\omega_m \frac{\omega_m \Lambda_2 \sigma(T) + \omega_a - i\alpha\omega}{\omega_\perp^2 - (1+\alpha^2)\omega^2 - 2i\alpha\omega_2\omega}.$$

The frequencies appearing in formulae (9) are defined by the expressions

$$\omega_m = \gamma p I_s, \qquad \omega_a = \gamma H_a,$$
$$\omega_\parallel^2 = (\omega_m \Lambda_1 \sigma(T) + \omega_a)(\omega_m \Lambda_3 \sigma(T) + \omega_a),$$
$$\omega_\perp^2 = (\omega_m \Lambda_2 \sigma(T) + \omega_a)(\omega_m [\Lambda_3 + \Lambda_4]\sigma(T) + \omega_a), \qquad (10)$$
$$\omega_1 = \omega_m (\Lambda_1 + \Lambda_3)\sigma(T)/2 + \omega_a,$$
$$\omega_2 = \omega_m (\Lambda_2 + \Lambda_4)\sigma(T)/2 + \omega_a,$$

The values of $\Lambda_1$, $\Lambda_2$, $\Lambda_3$ and $\Lambda_4$ are the following linear combinations of the tensor components

$$\Lambda_1 = (\Lambda_{F,zz} - \Lambda_{F,xx} + \Lambda_{AF,xx} - \Lambda_{AF,zz})/2 = 8.03,$$
$$\Lambda_2 = (\Lambda_{F,yy} - \Lambda_{F,xx} + \Lambda_{AF,xx} - \Lambda_{AF,yy})/2 = 8.03,$$
$$\Lambda_3 = (\Lambda_{F,yy} - \Lambda_{F,xx} + \Lambda_{AF,yy} + \Lambda_{AF,xx})/2 = 1.17, \qquad (11)$$
$$\Lambda_4 = (\Lambda_{F,zz} - \Lambda_{F,yy} + \Lambda_{AF,zz} - \Lambda_{AF,yy})/2 = 12.57.$$

Note that the value $\Lambda_4$ coincides with demagnetization factor $4\pi$ of a thin film within the accuracy of numerical calculation of the sums in (6). This result is natural since the grains lattice has a final number of layers along axis $z$ and infinite number of layers in other directions, that is, the lattice has demagnetization factor $4\pi$.

As it was mentioned above, the mean-field approach can be used to find the factor $\sigma(T)$ in expressions (9) and (10). Using corresponding partition functions, we can write the expression for the thermodynamic average for sub-lattice magnetizations at $T > 0$ in the form [18],

$$\langle \mathbf{M}_k(T) \rangle = \frac{\int d\mathbf{n}_k \cdot \mathbf{M}_k \exp\left\{ \left( \mathbf{M}_k \cdot \mathbf{H}_k^{\text{eff}} + \frac{H_a}{pI_s}(M_{k,x})^2 \right) \cdot \frac{V_f}{k_B T} \right\}}{\int d\mathbf{n}_k \cdot \exp\left\{ \left( \mathbf{M}_k \cdot \mathbf{H}_k^{\text{eff}} + \frac{H_a}{pI_s}(M_{k,x})^2 \right) \cdot \frac{V_f}{k_B T} \right\}}, \qquad (12)$$

where the integration is performed over directions of sub-lattice magnetic moments and an effective fields $\mathbf{H}_k^{\text{eff}}$ are defined by the following expression,

$$\mathbf{H}_{1,2}^{\text{eff}} = -\Lambda_{AF} \langle \mathbf{M}_{2,1}(T) \rangle - \Lambda_F \langle \mathbf{M}_{1,2}(T) \rangle. \qquad (13)$$

Equations system (12) can be reduced to a single equation. It is evident that $\langle \mathbf{M}_1(T) \rangle = -\langle \mathbf{M}_2(T) \rangle = \langle M(T) \rangle \mathbf{e}_1$. As a result, simple transformation of (12) gives rise to the equation for the parameter $\sigma(T)$ in the form,

$$\sigma(T) = \frac{\int_{-1}^{1} x \exp\left(x p I_s^2 V_f \Lambda \sigma(T)/4k_B T + x^2 I_s H_a V_f / 2k_B T\right) dx}{\int_{-1}^{1} \exp\left(x p I_s^2 V_f \Lambda \sigma(T)/4k_B T + x^2 I_s H_a V_f / 2k_B T\right) dx}, \quad (14)$$

where $\Lambda = \Lambda_{AF,xx} - \Lambda_{F,xx}$. Equation (14) allows one, in particular, to calculate the temperature $T_N$ of the destruction of the magnetic ordering in the system of ferromagnetic particles. It is determined by the condition $\sigma(T) = 0$ and the value $T_N$ could be considered as analog of the Neel temperature.

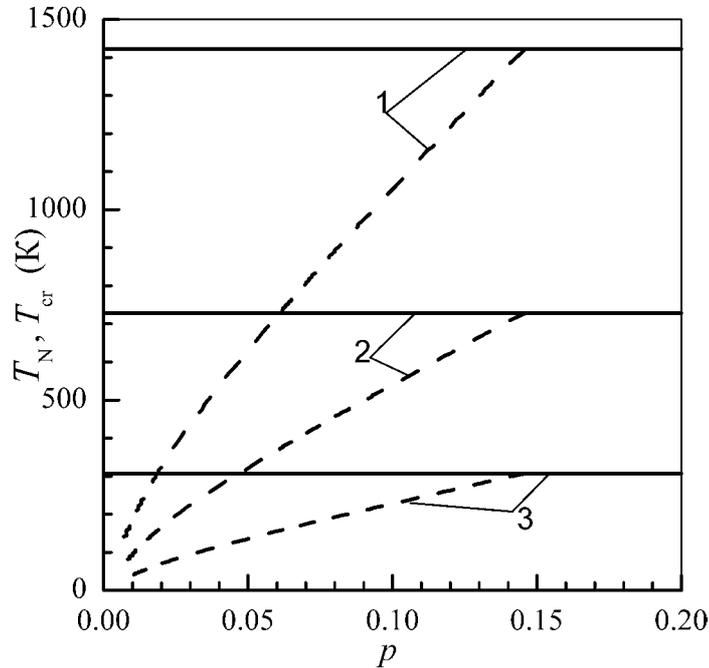

**Fig. 2.** Dependencies of $T_N$ (dashed lines) and $T_{cr}$ (solid lines) on the magnetic particle content $p$ at $I_s = 1500$ Gs, $H_a = 500$ Oe and different values of the particle diameters $D$: $1 - D = 10$ nm; $2 - D = 8$ nm; $3 - D = 6$ nm.

The above approach is obviously true only at temperatures much lower than the Curie temperature of the ferromagnetic grains. At $p = 0.2$, $I_s = 1500$ Gs, $H_a = 100$ Oe and particles diameter $D = 10$ nm, we estimate from equation (14) $T_N \approx 1700$ K and $\sigma(300\,\text{K}) \approx 0.944$. However, at lower density and/or for smaller particles the Neel temperature $T_N$ could be lower than the room temperature. For example, at $p = 0.05$, $I_s = 1500$ Gs, $H_a = 400$ Oe, and $D = 7$ nm we get $T_N \approx 240$ K. Since $\sigma(T > T_N) = 0$, formulae (9) and (10) for permeability components $\mu_{yy}$ and $\mu_{zz}$ of a composite film at high temperature are reduced to

$$\mu_{yy} = \mu_{zz} = 1 + 4\pi\omega_m \frac{\omega_a - i\alpha\omega}{\omega_a^2 - (1+\alpha^2)\omega^2 - 2i\alpha\omega_a\omega}, \quad (15)$$

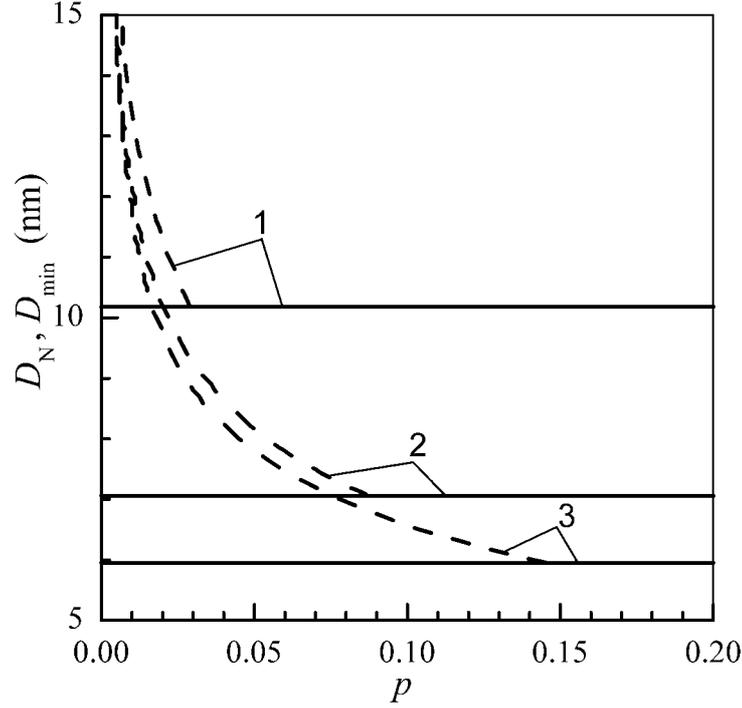

**Fig. 3.** Dependencies of $D_N$ (dashed lines) and $D_{min}$ (solid lines) on the magnetic particle content $p$ at $I_s = 1500$ Gs, $T = 300$ K and different values of the anisotropy field $H_a$: $1 - H_a = 100$ Oe; $2 - H_a = 300$ Oe; $3 - H_a = 500$ Oe.

At temperatures higher than transition temperature $T_N$ the composite is in a magnetically disordered (paramagnetic) state. In such a case the interaction between particles do not affect the composite permeability and expression (15) naturally coincides with the result for permeability of an array of non-interacting particles with parallel easy axes found by solving corresponding Landau-Lifshits-Gilbert equation (see, e.g., [17]).

The dependence of the antiferromagnetic order destruction temperature $T_N$ on the ferromagnetic particle volume fraction $p$ is shown in Fig. 2 at different particle diameters by dashed lines. The temperature $T_N$ naturally increases with the increase of the ferromagnetic particles density and diameters. At $T > T_N(p)$ the film permeability is the same as for non-interacting particles and at $T < T_N(p)$ the magnetic interaction between particles should be taken into account. If the diameter of the ferromagnetic inclusions is small, $D < D_{min}$, then at calculating magnetic permeability one should also take into consideration thermal relaxation. Solid straight lines in Fig. 2 present corresponding values of temperature $T_{cr}$ defined by the condition $\lambda = 1$,

$$T_{cr} = \pi H_a I_s D^3 / 12 k_B . \tag{16}$$

If $T > T_{cr}$ the thermal superparamagnetic relaxation becomes of importance.

In Fig.3 on the coordinate plane $(D, p)$ solid straight lines indicate the values of the minimal diameter of the particles $D_{min}$ found from expression (1). At $D > D_{min}$ one can neglect the contribution of thermal relaxation to the film permeability. The transition temperature $T_N$ evidently depends on the particles diameter, the larger is the diameter, the higher is $T_N$. At given temperature, we can define a characteristic particle diameter $D_N$ by the condition $T_N(D=D_N) = T$. If $D < D_N$ the composite magnetic permeability is determined by expression (15) for non-interacting particles, if $D > D_N$ the magnetic interaction is of significance and the permeability tensor should be calculated using equations (9) – (11) and (14). Dashed lines in Fig. 3 show the dependence of the value $D_N$ on $p$ found from the solution of equation (14). The composite magnetic permeability is to a great extent affected by the granule anisotropy field $H_a$. To illustrate the fact, the functions $D_{min}(p)$ and $D_N(p)$ are presented at different values of $H_a$. Both these diameters decrease with the growth of the anisotropy field. It is well known that the field of the magnetic anisotropy of nanocomposites essentially depends on the non-magnetic matrix composition and manufacturing method due to surface effects at the boarder of grains and magnetic matrix [5,6,19]. As a result, the nanocomposite anisotropy field can exceed bulk sample anisotropy field by several times or even by an order of magnitude.

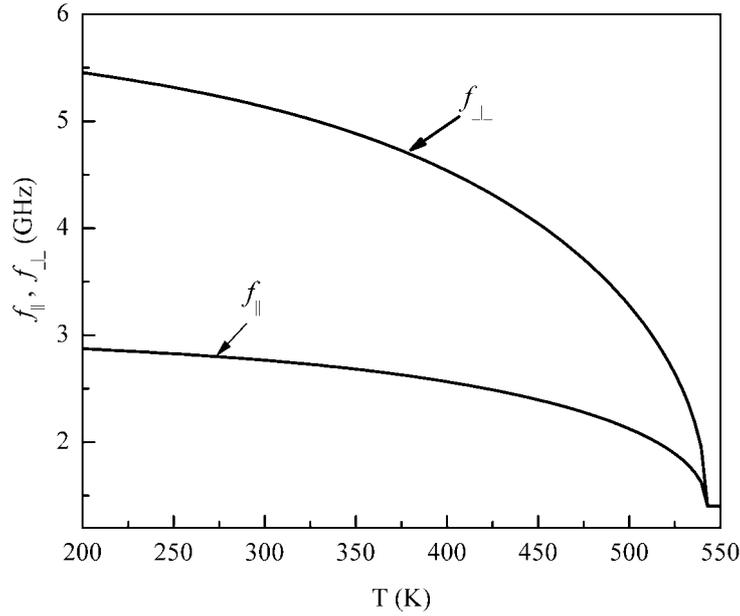

**Fig. 4.** Temperature dependence of the frequencies $f_\parallel$ and $f_\perp$ at $p = 0.1$, $I_s = 1500$ Gs, $H_a = 500$ Oe, and $D = 8$ nm.

If the damping parameter $\alpha$ is small, the frequencies $f_\parallel = \omega_\parallel / 2\pi$ and $f_\perp = \omega_\perp / 2\pi$ are resonance frequencies of magnetization rotation along the corresponding directions. As follows from expression (10), the resonance frequencies depend on the temperature, ferromagnetic grains density, magnetization, and anisotropy field. The resonance frequencies decrease monotonously with decrease of the grains volume fraction $p$. At low $p$ when the interaction between particles becomes insignificant, these two frequencies coincide and are equal to $\omega_a / 2\pi$. The calculations by formulae (10) at $I_s = 1500$ Gs, $p = 0.1 – 0.5$, and $H_a = 0.1 – 1$ kOe give that the frequency $f_\parallel$ lies within the range 1 – 15 GHz while the resonance frequency of oscillations in the direction perpendicular to the film plane, $f_\perp$, is in the range of 4 – 30 GHz. Such resonance frequencies are several times higher than that in the case of non-interacting particles. The temperature dependencies of the frequencies $f_\parallel$ and $f_\perp$ are depicted in Fig. 4 at $p = 0.1$, $H_a = 500$ Oe, and $D = 8$nm. For this case, the transition temperature $T_N$ is about 530 K. The resonance frequencies drop down with $T$ due to decrease of the parameter $\sigma(T)$ characterizing the magnetic ordering.

The frequency dependencies of imaginary parts of permeability tensor components are shown in Figs. 5 at different content of ferromagnetic particles $p$ and in Figs. 6 at different temperatures. The corresponding real parts can be calculated using Kramers-Kroning relation.

Curves 1 in Figs. 5, 6 correspond to the parameters range in which the film permeability is equal to the magnetic permeability of the system of non-interacting particles. The results presented in Figs. 4 – 6 demonstrate that the interaction between ferromagnetic inclusions leads to a substantial increase of resonance frequencies. Note also that both the real $\mu'$ and imaginary $\mu''$ parts of the longitudinal component of permeability tensor $\mu_{yy}$ could be rather high ($\mu_{yy} \sim 10$) in the frequency range of several GHz.

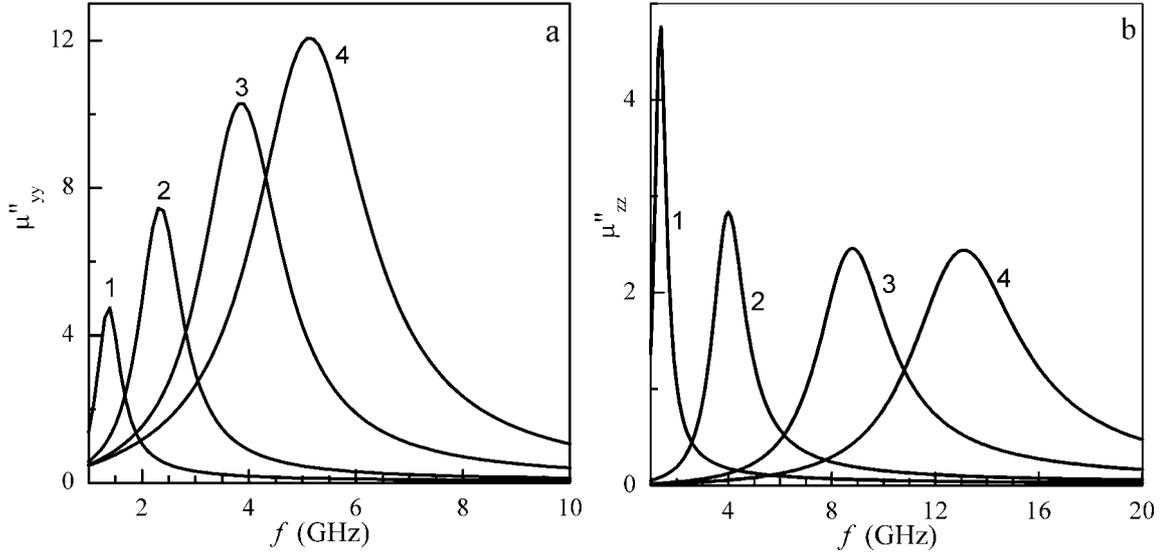

**Fig. 5.** Frequency dependence of the imaginary $\mu''$ parts of the longitudinal (a) and transverse (b) components of the magnetic permeability tensor at $I_s = 1500$ Gs, $H_a = 500$ Oe, $D = 10$ nm, $T = 300$ K, $\alpha = 0.2$ and different ferromagnetic particles content p: 1 – p = 0.05; 2 – p = 0.1; 3 – p = 0.2; 4 – p = 0.3.

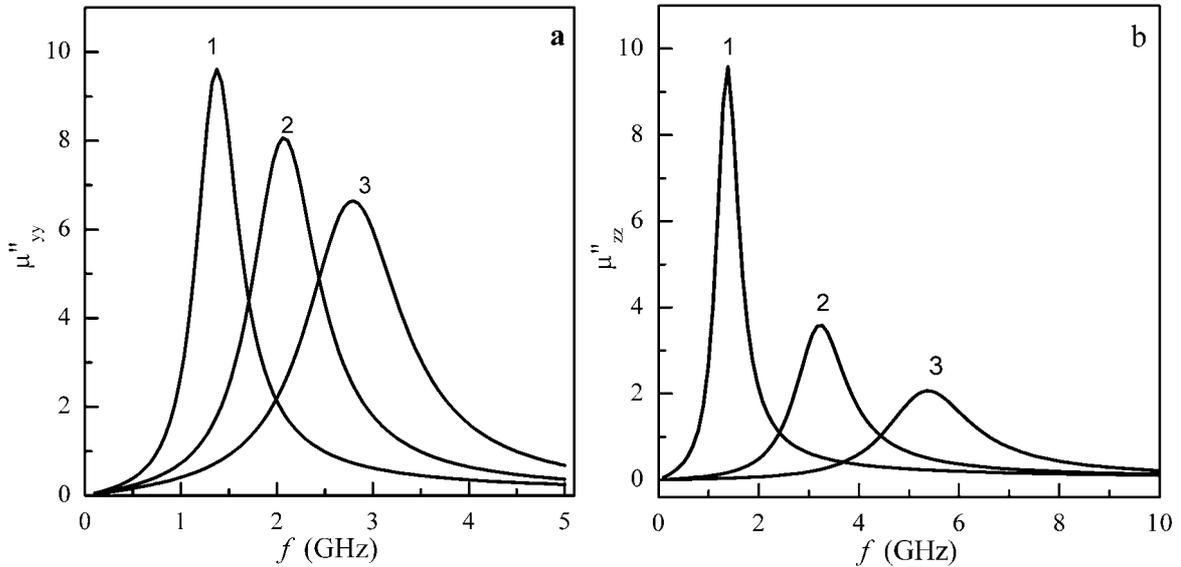

**Fig. 6.** Frequency dependence of the imaginary $\mu''$ parts of the longitudinal (a) and transverse (b) components of the magnetic permeability tensor at $p = 0.1$, $I_s = 1500$ Gs, $H_a = 500$ Oe, $D = 8$ nm and different temperatures: *1 – T = 600K; 2 – T = 500K; 3 – T = 200K.*

## 4. Conclusions

The present study is devoted to the high frequency magnetic permeability of composite films with ferromagnetic nanoparticles embedded into insulating non-magnetic matrix. It is supposed that the space between the ferromagnetic granules is much larger than the characteristic distance of exchange interaction between granules and the magnetic interaction between the particles is the magnetostatic dipole one. Let us discuss the main assumptions made in the paper. It is suggested a rather simple model of the ordered space distribution of the granules and ordered orientation of their magnetic moments. Within the framework of the considered model (a cubic lattice of spherical particles) the antiferromagnetic ordering of nanoparticles magnetization vectors turns out to be a ground state of the composite at low temperatures. The magnetic ordering noticeably affects the permeability and resonance frequencies at temperatures lower than the threshold (Neel) temperature $T_N$. It should be mentioned that under certain conditions a low-temperature equilibrium state of composites could be ferromagnetic. Low temperature ferromagnetic arrangements can appear due to either specific sample shape (a needle-like sample [15, 16]), or specific structure of the composite (particular triangular lattices of elongated particles [13, 14]). More complicated structures are also possible, for example, an antiferromagnetic arrangement of ferromagnetically ordered regions [7]. It seems natural to suppose that the cubic lattice and the antiferromagnetic arrangement should approximate properties of the disordered composite film more adequately than the specific structures listed above. The other essential assumption is the collinearity of the easy axis of the granules in the composite. However, this supposition seems natural in the case of the strong magnetic interaction between the granules since such a state has minimal magnetic energy. In addition, the film deposition in the magnetic field favors the formation of the collinear structures.

In the framework of our approach, the crossover between the ordered and disordered magnetic structures is a second order phase transition. It occurs in a stepwise manner at some temperature $T_N$. In any real composite the transition should be smeared due to some natural disorder in the positions, dimensions, and orientations of the ferromagnetic granules.

A complete and rigorous solution to the problem of the magnetic properties of the composite with strongly interacting particles is not yet found. The calculation of the small perturbations to the solution for the case of non-interacting particles could not be considered as a constructive approach to this problem. Our study corresponds to just an opposite limit, that is, the case of the system with a very strong magnetic dipole interaction. The composite films with the appropriate parameters are described in many studies. However, their high frequency properties are not studied up to now. We hope that the present paper will be incentive for such measurements.

Note, a rather high value of the magnetic permeability was reported in [3] for nanocomposite Co-Zr-O films within the frequency range up to several GHz. However, we suggest that the exchange interaction prevails in these systems. The high concentration of the ferromagnetic phase, small sizes of the magnetic granules, and relatively high conductivity of the composite (exceeding the Mott limit, 250 1/Ohm·cm, for a metal) testify to such a conclusion.

Let us discuss the obtained results. We have studied the case ($KV_f > k_B T$ or $\lambda > 1$) when the oscillations of the magnetic moments of the granules have a resonance nature and the magnetic permeability of the composite is determined by the rotation of the ferromagnetic inclusions magnetic moments near their equilibrium directions in high frequency magnetic fields. The calculations demonstrate that the contribution of this mechanism to the nanocomposite permeability dominates in a rather wide range of the sizes and densities of the magnetic nanoparticles. In the limit of non-interacting magnetic granules, the ferromagnetic resonance frequency is determined only by the anisotropy field of the particles and is equal to $\omega_a = \gamma H_a$.

The magnetic dipole interaction of the particles leads to a shift of the resonance frequencies towards higher values. In this case, the resonance frequency depends on the ferromagnetic phase content, magnetic particle sizes, temperature, and the sample geometry as well. The composite film permeability can achieve rather high values, which is quite promising for high-frequency applications of nanocomposites.

The thermoactivated re-orientation of the ferromagnetic particles magnetic moments becomes of significance in the parameter range where $\lambda<1$. In this case the behavior of the granule magnetic moment is similar to the motion of the Brownian particle and can be described by the Focker-Plank equation (see, e.g. [11,12,20]). Nevertheless even at $\lambda<1$ the model used in this paper remains valid in the temperature range $T<T_N$ where the dipole interaction between the magnetic particles is of importance. Really, at low temperatures the internal effective magnetic field produced by the ferromagnetic particles prevents the thermal motion of the granule magnetic moments. This effect should give rise to the decrease of the absorption of energy with the decrease of temperature in the frequency range, in which the thermal relaxation mechanism dominates at higher temperatures. Such an effect was observed in the experiment [7].

A more detailed analysis of the high frequency magnetic permeability of nanocomposite films requires taking into account effects of a disorder in orientations of nanoparticles magnetic easy axes, deviations of magnetic grains shape from spherical one, and distribution of particles by their sizes observed in real nanocomposite materials (e.g. [19]). Besides, the high frequency magnetic permeability could be affected by disorder in particles positions. Having the solution to the problem in the limit of the strong dipole interaction one could calculate the correction to it accounting the listed above factors. In particular, we intend further to use a composite model in which small deviations of magnetic grains positions from the points of the cubic lattice are allowed.

**Acknowledgements**

The study was performed with the support of the Russian Foundation for Basic Research (RFBR No 03-02-06320) and the Russian Federation Grant "Leading Scientific Schools" Н -1694.2003.2.